\documentclass[graybox,natbib]{svmult}

\usepackage{mathptmx}       % selects Times Roman as basic font
\usepackage{helvet}         % selects Helvetica as sans-serif font
\usepackage{courier}        % selects Courier as typewriter font
\usepackage{type1cm}        % activate if the above 3 fonts are
\usepackage{amssymb}
\usepackage{amsmath}
\usepackage{makeidx}         % allows index generation
\usepackage{graphicx}        % standard LaTeX graphics tool
\usepackage{multicol}        % used for the two-column index
\usepackage[bottom]{footmisc}% places footnotes at page bottom
\usepackage{natbib}
\makeindex             % used for the subject index
\newcommand{\msun}{\ensuremath{\mathrm{M}_\odot}}
\newcommand{\mch}{\ensuremath{M_{ch}}}
\newcommand{\nuc}[2]{\ensuremath{\mathrm{^{#1}#2}}}
\begin{document}

\title*{Nucleosynthesis in thermonuclear supernovae}
\author{Ivo Seitenzahl and Dean Townsley}
\institute{Ivo Rolf Seitenzahl \at Research School of Astronomy and Astrophysics, Australian National University, Canberra, ACT 2611, Australia \email{ivoseitenzahl@gmail.com}
\and Dean Martin Townsley \at Department of Physics and Astronomy, The University of Alabama, Tuscaloosa, AL 35487-0324 \email{dean.m.townsley@ua.edu}}
\maketitle

\abstract{The explosion energy of thermonuclear (Type Ia) supernovae is derived from the difference in nuclear binding energy liberated in the explosive fusion of light ``fuel'' nuclei, predominantly carbon and oxygen, into more tightly bound nuclear ``ash'' dominated by iron and silicon group elements.
The very same explosive thermonuclear fusion event is also one of the major processes contributing to the nucleosynthesis of the heavy elements, in particular the iron-group elements.
For example, most of the iron and manganese in the Sun and its planetary system was produced in thermonuclear supernovae.
Here, we review the physics of explosive thermonuclear burning in carbon-oxygen white dwarf material and the methodologies utilized in calculating predicted nucleosynthesis from hydrodynamic explosion models.
While the dominant explosion scenario remains unclear, many aspects of the nuclear combustion and nucleosynthesis are common to all models and must occur in some form in order to produce the observed yields.
We summarize the predicted nucleosynthetic yields for existing explosions models, placing particular emphasis on characteristic differences in the nucleosynthetic signatures of the different suggested scenarios leading to Type Ia supernovae.
Following this, we discuss how these signatures compare with observations of several individual supernovae, remnants, and the composition of material in our Galaxy and galaxy clusters.
}

\section{Introduction}
\label{intro}
``Thermonuclear supernovae" \index{Thermonuclear supernovae} is the collective name given to the family of supernova explosions that derive the energy that drives their expansion against gravity from the exothermic transmutation of less tightly bound nuclear ``fuel" (such as e.g.  \nuc{4}{He}, \nuc{12}{C}, or \nuc{16}{O}) into more tightly bound nuclear ``ashes'' (such as e.g. \nuc{28}{Si} or \nuc{56}{Ni}). This is in contrast to ``core-collapse supernovae", which, in spite of a contribution to the explosion energy by nuclear burning behind the outgoing shock-wave, are largely powered by the gravitational binding energy that is released during the stellar core collapse. A key difference between the thermonuclear burning in thermonuclear supernovae and core-collapse supernovae and is that the nucleosynthesis in the former generally proceeds at higher fuel density and therefore at lower entropy.  For a (text) book on supernovae and nucleosynthesis see \citet{arnett1996a}.

Although details of the progenitor system and explosion mechanism are still unknown, there is a general consensus that thermonuclear supernovae are the physical mechanism behind what are observationally classified as ``Type Ia supernovae" (SNe Ia); the two terms are therefore often used interchangeably. Note that while it is likely that all SNe~Ia are of thermonuclear origin, the converse need not be true.
A key aspect all sub-classes of thermonuclear supernova models have in common is that they all feature explosive (meaning dynamical) thermonuclear burning in white dwarfs (WD), stars that are supported against gravitation collapse by electron degeneracy pressure.  

We begin by summarizing the physical processes of explosive fusion involved in thermonuclear supernovae and how the interplay of these with the structure and dynamics of the exploding WD lead to the major yields of the explosion.
A short discussion of the techniques used to compute yields in modern simulations follows.
Our discussion of the expectations for the yields of thermonuclear supernovae then proceeds by treating several of the major proposed scenarios for the explosion.
The nucleosynthesis of p-nuclei are discussed separately, and we end the chapter with a discussion of the observational signatures of various aspects of nucleosynthesis in thermonuclear supernovae.

\section{Physics background and methodology}
\label{s0}
\subsection{Explosive thermonuclear burning}

The key fundamental aspect of nucleosynthesis in a Type Ia supernova comes from the realization, established by \citet{nomoto1984a},
that if a near-Chandrasekhar mass carbon-oxygen (CO) WD star is incinerated on a similar timescale to its dynamical time,
the resulting ejecta gives rise to a transient that is spectroscopically characteristic of a Type Ia.
That is, it is poor in H and He and shows strong Si features at maximum light,
with a power source interior to the maximum light photosphere provided by the radioactive decay of $^{56}$Ni.
Exactly how such an incineration of a WD comes about,
and even where the massive CO WD required by some scenarios might come from are still mysterious, but the basic model remains clear.

\begin{figure}[b]
\sidecaption
\includegraphics[width=0.62\textwidth]{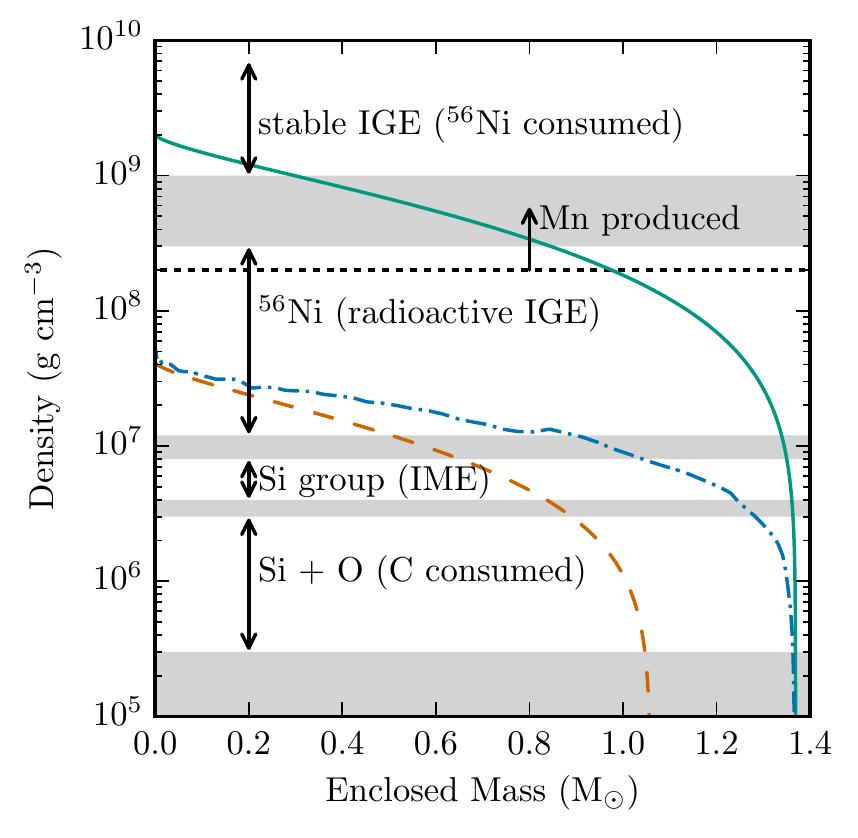}
\caption{
Nucleosynthetic products are determined principally by the density of the fuel when the reaction front passes.
High densities lead to more complete burning.
Profiles of WDs of several masses are shown,
demonstrating that a WD close to the Chandrasekhar mass (solid green), if burned without any expansion, produces almost solely IGE, while a 1.05 M$_\odot$ WD (dashed orange), will not produce Mn or stable IGE due to electron capture.
In the delayed-detonation family of scenarios,
early burning expands the $M_{\rm Ch}$ WD to a state in which significant amounts of IME are produced upon full incineration (dot-dashed blue).
\label{fig:density_demo}
}
\end{figure}

Some time spent on why incineration of a WD leads to the observed synthesized ejecta is worthwhile.
The density structure of a hydrostatic WD, along with the density dependence of the nucleosynthetic outcome of explosive CO fusion,
lead naturally to an ejecta structure in which iron group elements (IGE), including $^{56}$Ni, make up the inner regions,
surrounded by Si-rich outer layers.
\index{Carbon Fusion, Explosive}
Figure \ref{fig:density_demo} shows roughly how the products of burning depend on the local fuel density,
with high densities leading to more complete processing to IGE in nuclear statistical equilibrium (NSE),
and lower densities truncating the nuclear processing with the production of Si-group intermediate mass elements (IME).
\index{Iron Group Elements}
\index{Intermediate Mass Elements}
In NSE, material is hot and dense enough that all nuclear reactions that preserve the number of protons and neutrons can be approximated as occuring quickly.
As a result, the composition is determined by the relative binding energies of various species and statistical mechanical considerations and does not depend on individual reaction rates or, in a computation, their uncertainty.

Also shown in Figure \ref{fig:density_demo} is the density structure of a near-Chandrasekhar mass ($M_{\rm Ch}$) WD (1.38 M$_\odot$), as well as a 1.05 M$_\odot$ WD.
It is immediately clear that the hydrostatic state of the WD at the time the burning takes place has a direct role in the nucleosynthetic yield produced.
A $M_{Ch}$ WD burned by a detonation, for example, with no opportunity to expand, will produce nearly all IGE.
Comparatively, a 1.05 M$_\odot$ star subjected to a detonation will produce only about 0.6 $M_\odot$ of IGE, with the balance being mostly IME,
all simply due to the lower overall density.
A similar balance of yields favoring more IME can be obtained if the beginning of the incineration occurs slightly more slowly than the dynamical time of the star (around 1 second).
Then the star can expand as shown in the middle curve in Figure \ref{fig:density_demo},
so that the lowered densities lead to the production of IME in the outer region.

We will only discuss the physics of the relevant modes of combustion briefly.
The deflagration, or flame, mode is more slowly propagating and therefore able to allow the star to globally respond and expand.
Its propagation is driven by heat conduction in the degenerate electron medium.
The faster mode, detonation, is propagated by a shock that is self-sustained by the energy release and propagates across the star in less than the dynamical time.

\index{density -- effect on CO burning products}
The observed nucleosynthesis of SNe~Ia shows evidence of a large fraction of stable IGE in the innermost regions of the ejecta
\citep[e.g.,][]{Mazzalietal2007,Mazzalietal2015}.
As seen in Figure \ref{fig:density_demo}, synthesis of stable IGE requires densities above a few $10^8$~g~cm$^{-3}$.
These densities are only present in CO WDs above about 1.2 M$_{\odot}$, which can only be formed by accretion.
\index{electron capture}
\index{neutron enrichment}
Conversion of protons to neutrons by capture of electrons proceeds at high densities because the Fermi energy of the degenerate electrons, which hold the star up against gravity, is high enough for this conversion to be favorable.
At densities above a few $10^8$~g~cm$^{-3}$ the conversion rate is fast enough that,
in less than the dynamical time of the expanding star, the $^{56}$Ni-dominated NSE is changed first into one in which $^{54}$Fe and $^{58}$Ni, both of which are stable, are the dominant species.
Longer exposure can lead to even more neutron-rich products.
Comparison of the balance of neutron-enriched and non-enriched isotopes produced by the explosion to those found in the solar distribution constrains the central density during the initial part of the explosion
\citep{brachwitz2000a}.
These observational constraints are important due to uncertainty in the ignition physics \cite[e.g.,][]{gasques2005a,gasques2007a}.
% Ivo, you mentioned  \nuc{48}{Ca} \citep{woosley1997a}, but that should be out-of-date because the weak rates changed with the above Brachwitz work.

In addition to its importance for the isotopic abundances, neutron enrichment also decreases the amount of radioactive $^{56}$Ni available to power the observed bright transient.
The overall neutron richness of the material is determined by both the amount of electron capture during the supernova, as just described, as well the initial metallicity and electron capture during the pre-explosion core convection phase.
The initial metallicity controls the initial neutron excess mostly via the abundance of $^{22}$Ne in the progenitor WD \citep{timmes2003a}.
Almost all the initial C, N, and O elements in the progenitor star are first converted to $^{14}$N in the H-burning CNO cycle.
Then during the He burning phase that forms the CO WD, $\alpha$ captures and beta decay convert this $^{14}$N into $^{22}$Ne.
Core ignition of a thermonuclear SN is preceded by a phase of convective core carbon burning.
During this phase, electron capture near the center increases neutron enrichment \citep{chamulak2008a,piro2008a,martinez-rodriguez2016a}, with significant uncertainty due to the convective Urca process
\citep{stein2006a}.
\index{simmering phase}
For sub-$M_{\rm Ch}$ scenarios, the settling of the $^{22}$Ne within the WD \citep{bildsten2001a} may be important for the distribution of stable IGE in the ejecta.
Metallicity can also change the explosion in other ways \citep{calder2013a} including modification of the DDT density \citep{jackson2010a}.

Another product only obtained in large quantities from high-density burning is manganese \citep{seitenzahl2013a}.
\index{manganese}
This is produced as $^{55}$Co in the explosion and then decays.
Figure \ref{fig:density_demo} shows the dividing line, at about $2\times 10^8$ g~cm$^{-3}$, between alpha-rich freeze-out of NSE at lower densities and normal NSE freeze-out at higher densities.
At high densities, and therefore comparatively lower entropies for similar energy deposited, as the temperature falls, freeze-out occurs as the reactions that maintain the NSE become starved of $\alpha$ and other light particles.
By comparison, somewhat lower densities produce entropies at which $\alpha$ and other light particles are still present during freeze-out and destroy some species present in the NSE, including $^{55}$Co.
The solar ratio of Mn to Fe is higher than that produced by core-collapse SNe, thus requiring production in thermonuclear SNe.
This therefore implies a significant fraction of events with material burned above this threshold density \citep{seitenzahl2013a}, another indication of near-$M_{\rm Ch}$ progenitors.

\subsection{Methodology}

Two aspects of the combustion length scales in SNe~Ia make it challenging to compute accurate yields in explosion simulations.
Both of these are related to the relatively small length and time scales on which the reactions involved occur \citep[For a summary see the introduction of][]{townsley2016a}.
While the WD is of order several $10^8$~cm across, the length scales of the combustion front can be microns for the highest densities to cm for moderate densities $\sim 10^7$ g cm$^{-3}$.
In addition to this, the contrast in reaction length and time scales within the reaction front can be similarly vast.
At $10^7$~g~cm$^{-3}$, the length scale for full conversion to IGE in a planar detonation is $\sim 10^8$~cm, while the peak Si abundance only occurs some $10^3$~cm behind the shock,
and the carbon consumption scale is of order cm \citep{seitenzahlmeakin2009a,townsley2016a}.
These two contrasts in length scale make it computationally infeasible to compute the reaction front structure explicitly in simulations of the explosion of the star.
\index{reaction length scales -- CO burning}

Another challenge comes from the number of species involved in the nuclear reactions.
Accurate computation of carbon combustion with a nuclear reaction network \citep{hix2006a} requires of order 200-300 species in order to accurately capture combustion, silicon burning, and finally the electron
capture at high densities \citep{calder2007a,seitenzahltownsley2009a}.
Larger reaction networks can give a few percent improvement in accuracy, but around 200-300 appears to be sufficient.
For most of the history of SNe~Ia simulations, it was not possible to perform star-scale fluid dynamics simulations with such a large number of species.
This is currently improving with current very large machines having a high degree of parallelism in each node,
but there will continue to be a trade-off between fidelity of nuclear processes with more species and resolution of hydrodynamic processes such as turbulence with more spatial cells.
\index{reaction network size}

\begin{figure}[b]
\sidecaption
\includegraphics[width=0.62\textwidth]{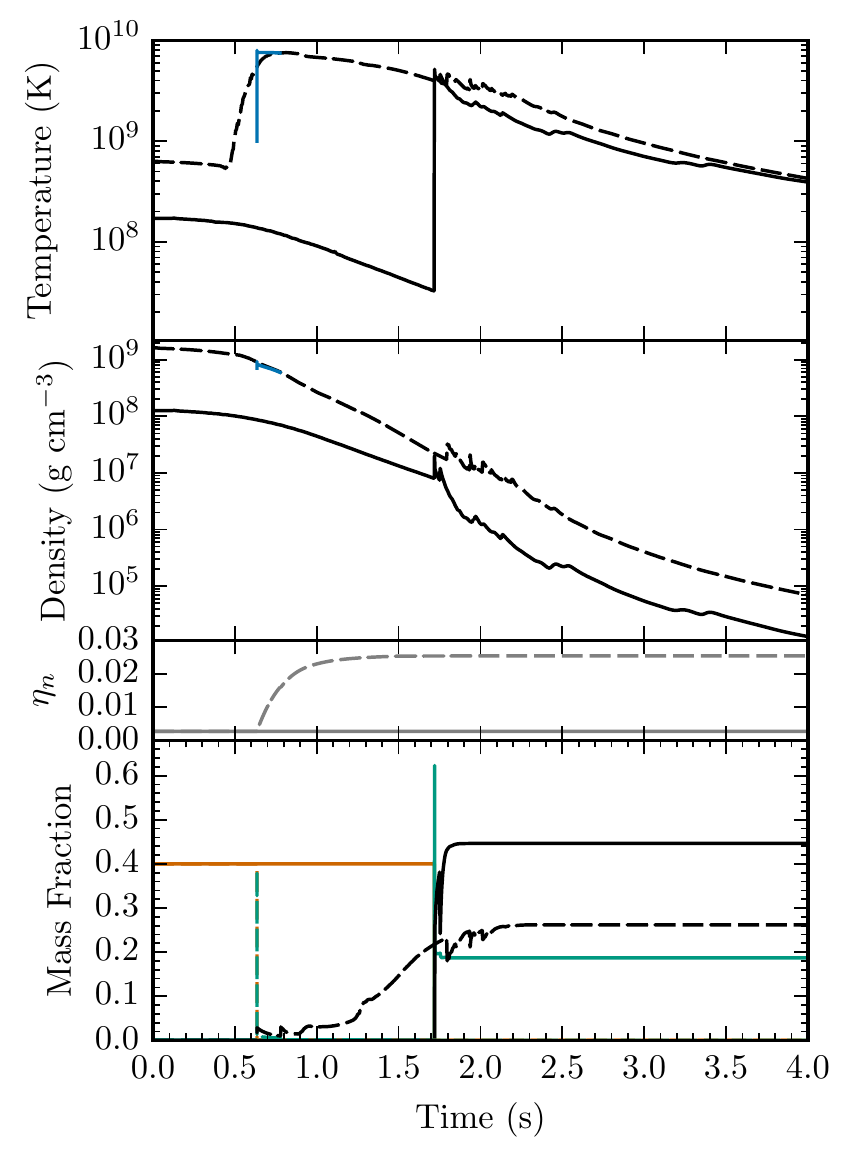}
\caption{
Nucleosynthesis is computed by processing the temperature-density history of a fluid element,
shown in the top two panels, computed in a hydrodynamic simulation of the explosion.  
The lower two panels show the resulting abundance neutron excess and abundance history.
The solid lines show a track processed by the detonation which only partially processes IME to IGE.
Only a few major abundances are shown: $^{12}$C (red), $^{28}$Si (green), and $^{56}$Ni (black).
The dashed lines show a track processed by the deflagration which has the $^{56}$Ni yield reduced by electron capture, enhancing the neutron excess $\eta_n$.
The short blue sections in the upper panels are the portion of the time history that is reconstructed \citep{townsley2016a}.
\label{fig:track_demo}
}
\end{figure}

\index{resolution}
These two challenges, of unresolved reaction length scales and the preference for good spatial resolution in multi-dimensional simulations, preferably 3D, instead of many species,
have led to the current strategy in which the reactions in the explosion
simulation are modeled in some way and then the yields are determined via post-processing of Lagrangian tracers \citep{travaglio2004a,seitenzahl2010a,townsley2016a}.
The post-processing proceeds by using Lagrangian fluid histories, often called ``tracks,'' ``trajectories,'' or ``tracers,'' recorded at many places in the hydrodynamic simulation of the explosion.
Two examples of the density and temperature history tracks recorded in this manner from a
2-dimensional simulation (assuming azimuthal symmetry) of the deflagration-detonation transition (DDT) scenario \citep{townsley2016a} are shown in Figure~\ref{fig:track_demo}.
The top two panels show tracks for fluid elements that were processed in the detonation (solid) and deflagration (dashed) modes.
Another advantage of separating the hydrodynamics and nucleosynthesis is that the influence of some changes, such as minor rates or initial abundances,
can be evaluated based on a smaller number of hydrodynamic simulations \citep[e.g.\ ][]{bravo2012a,parikh2013a,miles2016a}.
\index{post-processing}
\index{tracers}

This strategy of computing the explosion using large eddy simulation (LES) is neither unique to astrophysics nor to SNe~Ia.
In other contexts it may be used in circumstances where the physical turbulence dissipation scale is unresolved, but the behavior of turbulence in small scales is relatively well understood from the Kolmogorov model of turbulence and its successors.
\index{turbulence}
As a result, it is possible to do a valid simulation by only explicitly simulating the largest scales in an LES.
Supernovae present the additional complication of reactions, which are also subgrid scale like the turbulent dissipation.
The necessity of having a combustion model that accurately captures unresolved phenomena then introduces the topic of calibration and verification of that model.
The development of improved combustion models for LES and their verification is therefore a central topic in simulation of SNe~Ia and continues as increasingly detailed comparisons to observations are performed
\citep{schmidt2006b,seitenzahl2010a,ciaraldischoolmann2013a,jackson2014a,townsley2016a}.
In addition to uncertainty due to the modelling of unresolved processes in the hydrodynamics, there are also reaction rate uncertainties \citep{bravo2012a,parikh2013a}.

Examples of two tracks processed under different conditions in the same explosion, taken from \citet{townsley2016a}, are shown in Figure~\ref{fig:track_demo}.
The solid lines show the track for a fluid parcel that undergoes combustion via a detonation late in the simulation, about 1.7 s.
As can be seen, most of the nuclear processing of this parcel occurs near the reaction front or within about 0.2 s after the detonation passage.
The density when the detonation shock arrives is around $10^7$ g cm$^{-3}$, leading to incomplete burning of Si, as seen in the final abundances for this track, which have a mixture of IME and IGE.
No significant electron capture takes place so that the neutron excess $\eta_n = \sum_i (N_i-Z_i)Y_i= \sum_i(N_i+Z_i)Y_i -2\sum_iZ_iY_i=1-2Y_e$ is unchanged.
Here $Y_e$ is the number of electrons per baryon, $Y_i$ is the number of nuclei of species $i$ per baryon, $N_i$ and $Z_i$ are the number of neutrons and protons in the nucleus of species $i$, and charge balance is assumed.
\index{neutron excess}
The dashed lines show the track for a fluid parcel processed by the deflagration earlier in the simulation, about 0.6 s.
The deflagration track's final abundances are markedly different from the detonation, mostly because the chosen track shows a significant degree of conversion of protons to neutrons by electron capture, as show by the difference in $\eta_n$.
Immediately after passage of the deflagration front, all the fuel as well as the synthesized Si have been destroyed, and the $^{56}$Ni fraction is low due the high-temperature, high-density NSE state;
much of the material is in the form of $\alpha$ particles.
As the material cools and expands, the $^{56}$Ni abundance rises, but only to a fraction of about 0.3 by mass, because most of the IGE are in the form of more neutron rich species like $^{54}$Fe and $^{58}$Ni.
While both detonation and deflagration fronts are unresolved in this 4 km resolution simulation, in Figure~\ref{fig:track_demo} it is much more obvious for the deflagration case because the reaction front moves more slowly.
At the deflagration propagation speed, which is $\sim$km s$^{-1}$, thousands of times less than the sound speed,
it takes several tenths of a second for a fluid parcel to pass through the thickened reaction front that is a few times the grid resolution in thickness.
In order to improve yield accuracy in the context of this artificial broadening,
\citet{townsley2016a} developed a technique in which part of the thermal history recorded during the simulation is replaced in post-processing by a separate computation of the fully resolved structure of a steady-state deflagration (blue curve in upper panels).
The temperature rise in such a reconstruction is much faster, giving a more physically realistic temperature peak and therefore total integrated amount of electron captures.

\section{Explosive nucleosynthesis predictions of thermonuclear supernova models}
\label{models}
We give here only a broad overview that focuses on the qualitative differences of existing models in their respective nucleosynthetic yields. There are many different ways of categorizing thermonuclear SNe. From a nucleosynthesis point of view, it makes  most sense to group the explosion models into \textit{i)} near-\mch\ models (involving deflagrations ignited at high density and the associated neutronization) and \textit{ii)} pure detonation models of less massive WDs and \textit{iii)} models including detonations of thick He-shells and the special nucleosynthesis products thereof.   
\index{SNe Ia scenarios}

\subsection{Thermonuclear SNe from near-$M_{Ch}$ primary WDs}
\label{mch_models}
\index{near Chandrasekhar mass}
These models have in common the near central ignition of a deflagration via pycnonuclear fusion reactions as the accreting primary WD grows in mass and the central density increases. From a nucleosynthesis standpoint, these are generally the only models where in situ electron captures significantly lower the electron fraction and drive the yields towards more neutron-rich isotopes. Further, these are the only models where low entropy ``normal'' freeze-out from NSE occurs, with its typical nucleosynthetic signature, such as enhanced production of Mn (see discussion in Section \ref{observables}).
For these two reasons, deflagrations in high central density ($\rho \gtrsim 5 \times 10^9\,\mathrm{g}\,\mathrm{cm^{-3}}$) WDs have been suggested as the only plausible production site of certain neutron-rich intermediate mass and iron-group isotopes, such as \nuc{48}{Ca}, \nuc{50}{Ti}, or \nuc{54}{Cr} (see discussion in Section \ref{observables}). Since in-situ electron captures significantly increase the neutron-excess after ignition, the ignition density of the WD can have a greater effect on the final yields \citep[e.g.,][]{seitenzahl2011a} than the metallicity of the progenitor star \citep[e.g.,][]{miles2016a}. 

\subsubsection{Pure (turbulent) deflagrations}
\index{pure deflagration}
A detonation in a hydrostatic near-\mch\ CO WD \citep{arnett1969a} would burn most of the star at high density to \nuc{56}{Ni} and other IGE, producing insufficient IMEs such as Si and S to explain observed spectra of SNe~Ia. This has lead to deflagrations as the long-time favoured model, since a sub-sonic flame allows the WD to respond to the nuclear energy release with expansion to lower densities where the flame can produce IMEs. The widely referenced W7 model of a deflagration in a 1.38\,\msun\ CO WD \citep{nomoto1984a} has been very successful in reproducing key elements of the inferred structure of normal SNe~Ia, such as the mass of \nuc{56}{Ni} and the layered profile of the IMEs. However, W7 overproduced neutron rich nuclear species such as e.g., \nuc{54}{Cr}, \nuc{54}{Fe}, or \nuc{58}{Ni} \citep{thielemann1986a,iwamoto1999a}, a shortcoming that was ameliorated \citep{brachwitz2000a} when the electron capture rates of \citet{fuller1982a} were replaced with new electron capture rates for pf-shell nuclei from \citet{langanke2001a}. The initially slow evolution of the flame and the suppression of buoyancy still leads to enhanced electron captures in the W7 model even for the updated electron capture rates and hence too much \nuc{58}{Ni} \citep{maeda2010a}. The one-dimensional W7 model implements a sub-sonically moving flame that moves outward in the mass-coordinate, which means buoyancy and turbulent mixing are suppressed. As an unphysical consequence, the flame overruns the whole star and the burning ashes do not mix, retaining their original positions in the mass-coordinate. 
Three-dimensional pure deflagration models, which take buoyancy, fluid instabilities, and mixing into account, are too faint and chemically mixed to explain normal SNe~Ia \citep[e.g.,][]{roepke2007a} and their colors and spectra do not agree \citep{fink2014a}. Weakly ignited pure deflagration models that only partially unbind the star and leave a bound remnant behind do, however, provide excellent models for sub-luminous SNe Iax \citep{kromer2013a,kromer2015a}. While the ejecta of deflagration models that fail to completely unbind the whole WD \citep{fink2014a} are rich in IGE and contain little unburned \nuc{12}{C} and \nuc{16}{O}, more strongly ignited deflagrations that unbind the whole WD, such as e.g. the N1600 model from \citet{fink2014a}, eject $> 0.3\,\msun$ of both \nuc{12}{C} and \nuc{16}{O}, in addition to large amounts ($0.1\,\msun$ of \nuc{54}{Fe} and $0.07\,\msun$ of \nuc{58}{Ni}) of neutron-rich stable iron group isotopes, as well as $[\mathrm{Mn}/\mathrm{Fe}] > 0$.
Here we use the widespread ``bracket" notation: the logarithm of the ratio of element A to element B relative to the corresponding ratio in the Sun, $[A/B]=\log_{10}(A/B)-\log_{10}(A/B)_\odot$.

\subsubsection{Deflagration to detonation transition (DDT) models}
\index{Deflagration-Detonation Transition (DDT)}
To overcome the nucleosynthetic shortcomings of pure detonation and pure deflagration models, \citet{khokhlov1991a} introduced a delayed-detonation model, where an initial the sub-sonic deflagration
can transition to a super-sonic detonation under suitable conditions.
The expansion of the star prior to the detonation results in the desired nuclear burning at densities ${\lesssim}10^{7}\,\mathrm{g}\,\mathrm{cm^{-3}}$ where IMEs such as Si and S are synthesized. The significant contribution of the deflagration to IGE nucleosynthesis however still leaves an imprint on the overall yields, with overall $[\mathrm{Mn}/\mathrm{Fe}] > 0$ and (slightly) super-solar \nuc{54}{Fe}/\nuc{56}{Fe} and \nuc{58}{Ni}/\nuc{56}{Fe} \citep{seitenzahl2013b}. The white dwarf's carbon fraction (or C/O ratio) is only a secondary parameter; it does not influence the nucleosythesis in \mch\ models a lot since a lot of material burns into NSE \citep{ohlmann2014a}. DDT models tend to produce low light $\alpha$-element to Fe ratios, with typical O yields around $0.1\,\msun$ and Ne and Mg yields around $0.01\,\msun$ or lower.

\subsubsection{GCD and PRD models}
\index{Gravitationally Confined Detonation (GCD)}
The so-called gravitationally confined detonation (GCD) model \citep{plewa2004a} is based on a deflagration that ignites off-center in a single bubble and a detonation that is triggered just under the WD surface on the opposite side of the ignition point, when the deflagration ash compresses unburned fuel there after rising towards the surface and expanding laterally around the star. For this type of explosion model, detailed nucleosynthesis calculations that go beyond a basic 13 isotope $\alpha$-network have been presented for 2D simulations by \citet{meakin2009a} and for 3D simulations by \citet{seitenzahl2016a}. A key signature of GCD models is that although it is technically a near-\mch\ explosion model, very little mass is burned in the deflagration. As such, their isotopic nucleosynthesis signature resembles more that of pure detonations in massive sub-Chandrasekhar mass WDs, such as  presented in \citet{marquardt2015a}. [Mn/Fe] is sub-solar and products of neutronization, in particular the stable iron-group isotopes \nuc{54}{Fe} and \nuc{58}{Ni}, are not not overly abundant. Burning is quite complete, only ${\sim}0.1\,\msun\ \nuc{16}{O}$ and a few $\times 10^{-2}\,\msun$ of \nuc{12}{C} survive the explosion. 

\index{Pulsating Reverse Detonation (PRD)}
The pulsating reverse detonation (PRD) model \citep{bravo2009a} evolves initially analogous to the GCD model from a weak deflagration, although here the detonation is though to be triggered when an accretion shock forms during the contraction phase of the WD after the first radial pulsation brought about by the energy released in the deflagration. Since again not very much mass, 0.14 -- 0.26 \msun\ in the \citet{bravo2009b} models, is burned in the deflagration, the nucleosynthesis products do not carry the typical signature of near-\mch\ models, for example, [Mn/Fe] is sub-solar \citep{bravo2009b}.

\subsection{Thermonuclear SNe from detonating sub-M$_{Ch}$ WDs}
\label{submch_models}
Next, we group and discuss explosion models that are fundamentally based on detonations in sub-\mch\ WDs. Compared to near-\mch\ models, the smaller mass of the detonating primaries means that explosive nuclear burning at densities above ${\sim} 10^8 \mathrm{g}\,\mathrm{cm}^{-3}$ does generally not contribute to the yields. Consequently, the typical nucleosynthetic signatures of deflagrations, such as $[\mathrm{Mn}/\mathrm{Fe}] > 0$ or copious production of electron capture nuclei  \nuc{58}{Ni} or \nuc{54}{Fe}, are collectively absent in this group of models.
In the following, we briefly discuss the main proposed scenarios of detonating sub-\mch\ WDs, once again with a focus on their nucleosynthetic signatures.  

\subsubsection{Violent WD+WD merger}
\index{violent merger}
For a mass ratio close to unity, the \textit{violent merger} of two WDs is thought to provide a means to detonate at least one of the two WDs to produce a thermonuclear SN. 
The mass of the primary (heavier) WD is hereby the most important factor in determining whether the SN is sub-luminous \citep{pakmor2010a}, of ``normal" SN~Ia luminosity \citep{pakmor2012a}, or over-luminous \citep{moll2014a}, since in the violent merger models the primary WD detonates close to hydrostatic equilibrium \citep[cf.][]{sim2010a,ruiter2013a}. A small layer of helium present on the primary is thought to be essential in achieving robust detonations \citep{pakmor2013a}. Generally, the densities achieved in violent mergers are too low for low entropy freeze-out from NSE or enhanced production of neutron-rich electron capture nuclei (e.g., \nuc{54}{Fe} or \nuc{58}{Ni}) to occur. The isotopic composition of the IGEs therefore directly reflects the neutron excess and hence the progenitor system metallicity mainly through the abundance of $^{22}$Ne (see section \ref{s0}).

Whether or not the secondary WD detonates or gets disrupted is particularly important for setting the the overall production ratios, especially affecting the nucleosynthesis of the intermediate mass $\alpha$-elements, with a detonation in the lower mass secondary mostly (depending on the exact mass) ejecting \nuc{12}{C}, \nuc{16}{O}, \nuc{28}{Si}, and other IMEs \citep{pakmor2012a}. 

\subsubsection{Double-detonation models with He-shells}
\index{double detonation}
WDs accreting H at a low accretion rate $\lesssim \mathrm{few} \times 10^{-8}\,\msun \mathrm{yr}^{-1}$ process the H to He and accumulate thick He-shells that may ignite explosively \citep[e.g.,][]{taam1980a}. Similarly, such thick He-shells can be accumulated by directly accreting He, either from a degenerate He WD \citep{tutukov1996a} or a non-degenerate He-star \citep{iben1987a}. Multi-dimensional simulations have shown that detonations in the He-layer compress the core and likely (at least for high-mass CO cores) initiate a second CO detonation there  \citep[e.g.,][]{livne1990a,woosley1994a,shen2014a}, providing a possible explosion scenario for SNe~Ia.

\index{helium shell detonation}
Detonations in thick helium-shells of $\gtrsim 0.1\,\msun$ process enough material in the He-detonation to have a global impact on the yields; in particular $\alpha$-isotopes such as \nuc{36}{Ar}, \nuc{40}{Ca}, \nuc{44}{Ti} (decaying to stable \nuc{44}{Ca}), \nuc{48}{Cr} (decaying to stable \nuc{48}{Ti}), or \nuc{52}{Fe} (decaying to stable \nuc{52}{Cr}) are significantly enhanced compared to the products of explosive CO-burning \citep[e.g.,][]{woosley1994a,livne1995a}. The mass fraction of metals (e.g, C, N, O, Ne, ...) in the He-layer has a significant influence on the final nucleosynthesis products
\citep[e.g.,][]{kromer2010a,waldman2011a,moore2013a,shen2014b} and complicates model predictions.  

Double-detonation explosion models with relatively large primary WD masses $\gtrsim 0.9\,\msun$ and relatively small He-shell masses $< 0.1\,\msun$ provide an overall good match to observed ``normal" SNe~Ia, although current models still show some disagreement in the spectral and color evolution \citep[e.g.,][]{kromer2010a,fink2010a,woosley2011a}.
Systems with lower mass primary WDs have been proposed by \citet{waldman2011a} as potential progenitor systems of Ca-rich Gap transients \citep{kasliwal2012a} (SN~2005E-like events). \citet{waldman2011a} present detailed nucleosynthesis calculations for a range of low mass CO WD + He shell models. 

\subsubsection{Detonating ONe WDs}
\index{oxygen-neon white dwarfs}
Detonations may also be possible for scenarios where the primary star is an ONe WD \citep{marquardt2015a}. Compared to a generic CO primary WDs \citep[e.g.,][]{sim2010a}, the greater mass and hence compactness of the ONe WD results in an ejecta strongly dominated by Fe-group isotopes, with about 0.1\,\msun\ of \nuc{28}{Si} and $\lesssim 0.05\,\msun$ of \nuc{16}{O} ejected by the primary explosion \citep{marquardt2015a}. Naturally, very little carbon and direct products of carbon burning, such as Na or Ne, are ejected by detonating ONe WDs. 

\subsection{WD+WD collisions}
\index{collisions, WD}
\citet{papish2016a} present detailed isotopic yields for a range of CO WD + CO WD collision models, both with and without He-shells of various thickness. From a nucleosynthesis point of view, the important message is that the compression achieved during the collision (for the models investigated) is not enough to reach densities required for the copious production of electron capture nuclei or low-entropy freeze-out from NSE. The considerations for the nucleosynthesis of detonations in sub-\mch\ WDs and in He-shells therefore apply. 

\section{p-nuclei nucleosynthesis}
\label{pnuc}
\index{p-nuclei}
In the previous sections we have reviewed how explosive thermonuclear fusion of light nuclei such as \nuc{4}{He}, \nuc{12}{C}, and \nuc{16}{O} can synthesize heavier nuclei up to the iron-peak. The majority of the even heavier nuclides is synthesized by neutron capture processes, specifically the s- and r-processes \citep{cameron1957a,burbidge1957a}. A number of stable nuclides on the proton-rich side of the nuclear valley-of-stability however are ``excluded" \citep{cameron1957a} and cannot be synthesized by neutron capture processes as they are either bypassed or shielded from the s- and r-process paths by more neutron-rich stable isobars. The main p-nuclei are \nuc{74}{Se}, \nuc{78}{Kr}, \nuc{84}{Sr}, \nuc{92,94}{Mo}, \nuc{96,98}{Ru}, \nuc{102}{Pd}, \nuc{106,108}{Cd}, \nuc{113}{In}, \nuc{112,114}{Sn}, \nuc{120}{Te}, \nuc{124,126}{Xe}, \nuc{130,132}{Ba}, \nuc{138}{La}, \nuc{136,138}{Ce}, \nuc{144}{Sm}, \nuc{156,158}{Dy}, \nuc{162}{Er}, \nuc{168}{Yb}, \nuc{174}{Hf}, \nuc{180}{Ta}, \nuc{180}{W}, \nuc{184}{Os}, \nuc{190}{Pt}, and \nuc{196}{Hg}. p-nuclei contain less than one percent of the mass of nuclei heavier than $A=74$ in the Sun \citep{anders1989a}, yet they challenge our understanding of the synthesis of the heavy elements. The p-nuclei are thought to be synthesized via photo-disintegration ($\gamma$,n)-reactions of pre-existing s-process nuclei \citep[e.g.,][]{audouze1975a} in the  secondary $\gamma$-process \citep{woosley1978a}, requiring temperatures in the range $2.0\,\mathrm{GK} \lesssim T \lesssim 3.5\,\mathrm{GK}$; for reviews on p-nuclei and their nucleosynthesis see \citet{lambert1992a,pignatari2016a}. The nucleosynthesis site of the p-nuclei has been under debate for 60 years. Core-collapse supernovae (CCSNe) have long been favored \citep{arnould1976a}, but even modern calculations fail to reproduce the solar abundance pattern, most notably still underproducing the light p-nuclei $A\lesssim124$ \citep{rauscher2013a}. The origin of the \nuc{92,94}{Mo} and \nuc{96,98}{Ru} isotopes, which together comprise an unusually large 24\% and 7\% of the naturally occurring elements Mo and Ru respectively, has been particularly challenging to explain with CCSNe, due to a shortage of the relevant s-process seed nuclei \citep{woosley1978a}.
\index{molybdenum}
\index{ruthenium}

A further proposed site for p-nuclei nucleosynthesis are SNe~Ia \citep{howard1991a}. Scenarios where the progenitor system is in the single degenerate scenario are particularly promising, since the s-process seed abundance can be enhanced significantly in the thermal helium pulses during the accretion phase, with $\nuc{12}{C}(p,\gamma)\nuc{13}{N}(e^+\nu_e)\nuc{13}{C}(\alpha,n)\nuc{16}{O}$ as the main neutron source \citep{iben1981a}. This in situ enhancement facilitates a high p-nuclei yield in general agreement with solar abundances when the s-process seed is processed by the SN~Ia explosion \citep{travaglio2011a}. Whether CCSNe or SNe~Ia were the dominant sites of the nucleosynthesis of the p-nuclei in the Sun is still unsettled. However, SNe~Ia are currently the only viable proposed solution to the persistent \nuc{92,94}{Mo} and \nuc{96,98}{Ru} problem \citep{travaglio2011a}. Post-processing 2D delayed-detonation explosion model and performing Galactic chemical evolution calculations, \citep{travaglio2015a} find that more than 50\% of the p-nuclei (including Mo and Ru isotopes) can be synthesized by SNe~Ia, making SNe~Ia potentially the dominant site of the p-nuclei.

\section{Direct and indirect observable signatures of nucleosynthesis}
\label{observables}
\index{nucleosynthesis -- observable signatures}
In this final section we discuss the observable signatures of the nucleosynthesis occurring in thermonuclear supernovae. First, we summarize the current state of different approaches aimed at inferring isotopic (as opposed to elemental) production masses synthesized in the explosions. The most prominent isotope is \nuc{56}{Ni}, whose decay chain $\nuc{56}{Ni} \xrightarrow{6\,d} \nuc{56}{Co} \xrightarrow{77\,d} \nuc{56}{Fe}$ supplies most of the heat (in the form of interacting $\gamma$-rays and positrons) that make the ejecta of SNe~Ia shine brightly in optical light during the first ${\sim}3\,\mathrm{years}$ after the explosion. 
\index{cobalt decay}

\index{Arnett's rule}
The mass of \nuc{56}{Ni} synthesized in SNe~Ia can be approximately determined by \textbf{Arnett's rule:}
In its essence, Arnett's rule states that the bolometric luminosity of a SN~Ia at maximum light is proportional (with proportionality constant close to unity) to the instantaneous energy deposition rate by the radioactive decay \citep{arnett1982a}, which is dominated by the \nuc{56}{Ni} decay chain at that time. Thus, the \nuc{56}{Ni} mass of a SN~Ia can be directly estimated from the peak luminosity in a linear way. Although Arnett's rule is only approximate and contains some dependency on the explosion energy, opacity, total ejecta mass, or the spatial distribution of the \nuc{56}{Ni} \citep{pinto2000a}, it has been shown to yield \nuc{56}{Ni} masses that are consistent with those derived from more sophisticated radiative transfer modeling of the nebular spectra of SNe~Ia \citep{stritzinger2006a}. While Arnett's rule applied to normal SNe~Ia yields typical \nuc{56}{Ni} masses in the relatively narrow range $0.3\,\msun \lesssim M(\nuc{56}{Ni}) \lesssim 0.8\,\msun$ \citep{scalzo2014a}, the range covered by putative thermonuclear SN events including all sub-classes is much larger, spanning from significantly more than 1.4\,\msun, e.g., SN 2007if \citep{scalzo2010a}, to just a few times $10^{-3}\,\msun$, e.g., SN~2008ha \citep{foley2009a}. 

For a handful of events it has  been possible to more directly determine isotopic production masses of a few radionuclides by observing the characteristic $\gamma$- and X-rays emitted in their decays. We now discuss these special events briefly in turn.

\textbf{SN~2014J:}
\index{SN~2014J}
A mere 3.5~Mpc away, SN~2014J in M82 has been the closest Type Ia supernova since SN~1972E. Due to its proximity, the SPI gamma-ray spectrometer on the INTEGRAL satellite was able to directly detect for the first time the $\gamma$-ray lines from the decay of \nuc{56}{Ni} \citep{diehl2014a} and \nuc{56}{Co} \citep{churazov2014a} synthesized in a SN~Ia. The total mass of \nuc{56}{Ni} derived from the $\gamma$-ray observations is around $0.6\,\msun$ \citep[e.g.,][]{churazov2015a}, in good agreement with estimates from Arnett's rule \citep{dhawan2016a} and unambiguously confirming that the \nuc{56}{Ni} decay chain indeed powers the early light curves of normal SNe Ia.

\textbf{G1.9+0.3:}
\index{SNR G1.9+0.3}
G1.9+0.3 is the youngest known supernova remnant in our Galaxy, with a putative explosion date around 1900 \citep{carlton2011a}. From detections of the 4.1\,keV X-ray line of Sc with the Chandra satellite, \citet{borkowski2010a} infer a production mass of $1-7\times 10^{-5}\,\msun$ of \nuc{44}{Ti}, in good agreement with the theoretical predictions for the leading explosion models, but ruling out models with detonations of thick He-shells for this particular event. 

\index{Tycho's SNR}
\textbf{Tycho's SNR:} In 2014 \citet{troja2014a} reported an excess in their observations of Tycho's SNR in the 60--85\,keV band of the Burst Array Telescope (BAT) on the Swift satellite. However, their derived 3$\sigma$ range for the flux in the 78 keV line, $2.4\times10^{-6} < F_{78} < 2.6 \times 10^{−5}\,\mathrm{photons}\,\mathrm{cm}^{-2}\,\mathrm{s}^{-1}$ is not very restrictive, especially given the rather uncertain distance to Tycho's SNR. For a distance of $2.5\,\mathrm{kpc}$, their 3$\sigma$ range corresponds to $3 \times 10^{-5} \lesssim M(\nuc{44}{Ti}) \lesssim 3 \times 10^{-4}$, which is in agreement with most of the explosion models discussion in section~\ref{models}, but again ruling out the double-detonation model. In ${\sim}750\,\mathrm{ks}$ of NuSTAR observations \citet{lopez2015a} do not detect any sign of \nuc{44}{Ti} and they note that in order for their non-detection to be consistent with the Swift/BAT results, the Ti cannot be at low velocity in the inner 2'. To be compatible, they require moderate or high-velocity \nuc{44}{Ti} ($v \gtrsim 7,000\,\mathrm{km}\,\mathrm{s}^{-1}$) and/or that the \nuc{44}{Ti} is distributed across most of the extend of the SNR ($\gtrsim 3'$). 

In addition to inferring isotopic production masses from direct line detections, there are indirect methods that can be used to infer isotopic yields. Although much less \nuc{57}{Ni} than \nuc{56}{Ni} is synthesized in the explosion and \nuc{57}{Co} does not emit any positrons, \citet{seitenzahl2009a} pointed out that, due to the longer half-life of \nuc{57}{Co} compared to \nuc{56}{Co}, the $A=57$ decay chain $\nuc{57}{Ni} \xrightarrow{36\,h} \nuc{57}{Co} \xrightarrow{272\,d} \nuc{57}{Fe}$ would none-the-less become the dominant radioactive energy source, heating the ejecta with the kinetic energy of internal conversion and Auger electrons emitted in the decay of \nuc{57}{Co}. At even later times, \nuc{55}{Fe} (produced as \nuc{55}{Co}) will become the dominant source of  radioactive heating. Different explosion models predict different production ratios between the main radionuclides \nuc{56}{Ni}, \nuc{57}{Ni}, \nuc{55}{Co}, and \nuc{44}{Ti}. Knowledge of these production ratios, determined for example from modelling late-time observations of nearby SNe~Ia, can therefore be used to discriminate between competing explosion models \citep{roepke2012a}, even for comparable \nuc{56}{Ni} masses.  For two nearby SNe~Ia it has been possible to indirectly infer a production mass in the $A=57$ decay chain from the interaction of the decay radiation with the ejecta. 

\index{late-time light curves}
\index{nickel 57}
\index{SN~2011fe}
\textbf{SN 2011fe}, a low velocity gradient \citep[cf.][]{benetti2005a} normal SN~Ia that exploded a mere $6.4\,\mathrm{Mpc}$ away in a relatively un-obscured region in M101, has been followed to very late times \citep[e.g.,][]{kerzendorf2014a}.
Modeling the nebular spectrum from the epoch of 1034 days presented in \citet{taubenberger2015a}, \citet{fransson2015a} conclude that \nuc{57}{Co} is required to explain the observed flux level at ${\sim}1000\,\mathrm{d}$. Although they do not derive a best-fit production value for the mass of \nuc{57}{Ni}, they show that the assumed 1.5 times solar production ratio of \nuc{57}{Ni} to \nuc{56}{Ni} under-predicts the observed flux by a factor of ${\sim}2$, which indicates that the 57/56 ratio of SN~2011fe was likely even greater. 

\index{SN~2012cg}
\textbf{SN 2012cg}, also a normal SN~Ia, was a bit further (${\sim}15.2\,\mathrm{Mpc}$ in NGC 4424) away than SN~2011fe, but HST photometry was still possible out to 1055 days past maximum light. A \nuc{57}{Ni} to \nuc{56}{Ni} production ratio of roughly twice the corresponding solar value provides the best fit to the observed light curve \citep{graur2016a}. After SN~2011fe, this is the second normal SN~Ia that seems to require a \nuc{57}{Ni} to \nuc{56}{Ni} production ratio of $\gtrsim 2$ times solar. Since explosion models that involve exploding WDs near-\mch\ (such as delayed-detonations) predict greater 57/56 ratios than competing explosion models with significantly less massive primary WDs (such as the violent merger models or double-detonation models), \citet{graur2016a} conclude that a near-\mch\ progenitor is more likely for SN~2012cg.

\index{Galactic chemical evolution (GCE)}
As a major nucleosythesis site of heavy elements, thermonuclear supernovae also have a significant impact on \textbf{Galactic chemical evolution} (GCE); for a review on GCE see \citet{mcwilliam1997a}.
Notably, SNe~Ia are thought to have produced approximately two thirds of the Fe present in our Galaxy today \citep{dwek2016a}, the other third coming from CCSNe. CCSNe originate from massive stars and
explode rather promptly after star-formation, whereas SNe~Ia also arise in old stellar populations, which results in a ``delayed" enrichment of the interstellar medium by SNe~Ia. Any differences
between CCSNe and SNe~Ia in their nucleosynthetic yields ratios therefore result in gradients of those elemental abundance ratios as a function of metal enrichment, or [Fe/H].
For example, compared to CCSNe,  the production ratios of the light $\alpha$-elements (in particular O, Ne, Mg) relative to Fe are much lower in SNe~Ia, which explains the observed decrease of [$\alpha$/Fe] with increasing [Fe/H] for $[\mathrm{Fe}/\mathrm{H}] > -1$ \citep[e.g.,][]{timmes1995a,kobayashi2009a}.
In addition to Fe, SNe~Ia are also the dominant production sites of Cr and Mn, as well as contributing significantly, perhaps even the majority, to other iron group elements (such as Ti, Ni) and the heavier $\alpha$-elements, such as Si, S, Ar, Ca \citep[e.g.,][]{dePlaa2013a}. 

\index{manganese}
Of all these, the mono-isotopic Mn is of particular interest. [Mn/Fe] indicates that SNe~Ia from near-\mch\ WD primaries cannot be uncommon, since they are the only viable proposed nucleosynthesis site that predicts $[\mathrm{Mn}/\mathrm{Fe}] > 0$. The reason for this is that high density is required for the entropy to be low enough such that \nuc{55}{Co}, which decays to \nuc{55}{Mn} via \nuc{55}{Fe}, largely survives the freeze-out from NSE. This provides strong evidence that explosive nuclear burning at high density, $\rho \gtrsim 2 \times 10^8\,\mathrm{g}\,\mathrm{cm}^{-3}$, must have contributed significantly to the synthesis of Fe in the Galaxy \citep{seitenzahl2013a}. It is worth noting here that this strongly density/entropy dependent \nuc{55}{Co} yield can be used as a model discriminant for individual, nearby SNe via the $5.9\,\mathrm{keV}$ X-ray emission from the decay of \nuc{55}{Fe} \citep{seitenzahl2015a} or via the effect of radioactive heating by X-rays and Auger electrons from decaying \nuc{55}{Fe} \citep{roepke2012a}. Of interest in the context of GCE are further the chemical peculiarities of Local Group dwarf galaxies \citep{kobayashi2015a}, which could be explained by the nucleosynthetic contribution of SNe Iax, in particular pure deflagration models that fail to completely unbind the WD and leave remnants behind \citep[e.g.,][]{kromer2013a,kromer2015a}. 

\index{intracluster medium}
\textbf{X-ray spectroscopy of hot intra-cluster medium} (ICM) has emerged as one of the most promising ways to measure chemical abundances to constrain SN explosion models \citep[e.g.,][]{dupke2000a,dePlaa2007a}. The hot ICM is generally optically thin to X-rays and close to collisional ionization equilibrium \citep[see e.g.,][]{dopita2003a}, making abundance measurements relatively straightforward \citep[for a review see e.g.,][]{boehringer2010a}. The gravitational potential of large galaxy clusters is deep enough to retain the SN ejecta and the metal abundance of the ICM therefore directly reflects the integrated yields of all SNe (core collapse and thermonuclear) up to the present. Fits to the observed ICM abundances \citep{mernier2016a} indicate that to explain the large abundances of Ar and Ca (and perhaps Cr) \citep{dePlaa2013a,mernier2016b}, a further component has to be invoked, in addition to the contributions from normal SNe~Ia and CCSNe. Since no such contribution is required to explain the abundances in the Sun \citep{mernier2016b}, this points at very old stellar populations (long delay-time) giving rise to these explosions. A promising match to these requirements could be 1991bg-like SNe, a scenario that could simultaneously explain the origin and morphology of the Galactic $511\,\mathrm{keV}$ anti-matter annihilation line \citep{crocker2016a}. Moreover, once again based on the Mn abundance, it is disfavored that most SNe~Ia originate from detonating sub-\mch\ WDs \citep{mernier2016b}.

\index{supernova remnants}
\textbf{X-ray observations of supernova remnants} can provide further independent information about the chemical elements synthesized by SNe~Ia and constrain explosion scenarios for individual SNRs. For example, the high Mn/Fe and Ni/Fe ratios determined for SNR 3C\,397 from \textit{Suzaku} observations is indicative of neutronized material that could only be explained by an exploding massive near-\mch\ WD \citep{yamaguchi2015a}.

\index{calcium 48}
Last but not least, the nucleosynthesis origin of a few neutron-rich intermediate mass and iron group isotopes is very likely also linked to thermonuclear SNe, in particular for \nuc{48}{Ca}, but also \nuc{50}{Ti}, \nuc{54}{Cr}, and others \citep[e.g.,][]{woosley1997a}. \citet{meyer1996a} showed that nucleosynthesis of \nuc{48}{Ca} cannot primarily occur in CCSNe since the formation and subsequent survival of the \nuc{48}{Ca} quasi-equilibrium cluster \citep{bodansky1968a,woosley1973a} not only requires high neutron excess but also low entropy, such as is only obtained in explosive thermonuclear burning at the highest densities obtained in near-\mch\ SN~Ia explosions, for example from explosive thermonuclear burning in ONeMg WDs. Recent three-dimensional simulations by \citet{jones2016a} cast serious doubt on the canonical wisdom that near-\mch\ accreting ONeMg WDs collapse to neutron stars after central Ne and O-burning is ignited, opening the exciting possibility that such events may indeed be the primary nucleosynthesis site of \nuc{48}{Ca} and a few other neutron rich isotopes.

\begin{acknowledgement}
IRS was supported during this work by Australian Research Council Laureate Grant FL09921.
\end{acknowledgement}

\bibliographystyle{spbasic}
\bibliography{biblio}

\end{document}